\documentclass{article}
\usepackage{apalike}
\usepackage{graphicx}
\usepackage{times} 
\usepackage{graphicx}

\usepackage{amssymb}

\begin{document}



\begin{center}
{\Large\bf INTELLIGENT ENCODING AND ECONOMICAL\\ \vspace{2mm} COMMUNICATION  IN THE VISUAL STREAM} \vspace{5mm} \\
{\large Andr\'as L{\H o}rincz}

E\"otv\"os Lor\'and University, Budapest, Hungary \\ email: lorincz@inf.elte.hu

\end{center}

\begin{abstract}
The theory of computational complexity  is used to underpin a recent model of neocortical sensory processing. We argue
that encoding into reconstruction networks is appealing for communicating agents using Hebbian learning and working on
hard combinatorial problems, which are easy to verify. Computational definition of the concept of intelligence is
provided. Simulations illustrate the idea.
\end{abstract}

\section{Introduction}\label{s:intro}

A recent model of neocortical information processing developed a hierarchy of reconstruction networks subject to local
constraints \cite{lorincz02mystery}. Mapping to the entorhinal-hippocampal loop has been worked out in details
\cite{lorincz00parahippocampal}. Straightforward and falsifying predictions of the model concern the temporal
properties of the internal representation and the counteraction of delays of the reconstruction process. These
predictions have gained independent experimental support recently \cite{egorov02graded,henze02single}. The contribution
of the present work is to underpin the model by the theory of computational complexity (TCC) and to use TCC to ground
the concept of intelligence (CoI). We shall treat the resolution of the homunculus fallacy \cite{searle92rediscovery}
to highlight the concepts of the approach.

\section{Theoretical considerations}\label{s:control_rec}

In our view, the problem of encoding information in the neocortical sensory processing areas may not be detachable from
CoI. We assume that the wiring of neocortical sensory processing areas developed by evolution forms an ensemble of
economical intelligent agents and we pose the question: What needs to be communicated between intelligent computational
agents? The intriguing issue is that although (i) CoI has meaning for us and (ii) this meaning seems to be measurable
in practice, nevertheless, (iii) CoI has escaped mathematical definition. In turn, our task is twofold: we are to
provide a model of neocortical processing of sensory information \textit{and} a computational definition of
intelligence.

According to one view , intelligent agents learn by developing categories \cite{harnad03can}. For example,
mushroom-categories could be learned in two different ways: (1) by `sensorimotor toil', that is, by trial-and-error
learning with feedback from the consequences of errors, or (2) by communication, called `linguistic theft', that is, by
learning from overhearing the category described. Our point is that case (2) requires mental verification: Without
mental verification trial-by-error learning is still a necessity. In our model, verification shall play a central role
for constructing the subsystems, our agents.

Problem solving and verification can be related by TCC. From the point of view of communication, there are only two
basic types of computational tasks. The first, which can be called as not worth to communicate (non-WTC) type is either
easy to solve and easy to verify, or hard to solve and hard to verify. The other type is hard to solve but easy to
verify and, in turn, it is of WTC type. For non-WTC type problems communication is simply an overhead and as such, it
is not economical. On the other hand, WTC type problems -- according to TCC -- may have exponential gains if
communication and then verification is possible. As an example, consider the NP-hard Traveling Salesman problem. The
complexity of the problem is known to scale exponentially with the number of cities, whereas time of verification
scales linearly. Note that `recognition by components' \cite{biederman87recognition} is also a problem subject to
combinatorial explosion.

We conclude that economical communication occurs only for WTC-type problems. The intelligent `agents' (the subsystems)
are subject to the following constraints: (1) Subsystems sense different components or need to learn to separate
components of a combinatorial tasks, or both. (2) Information provided by the subsystems are to be joined by either
direct communication between them, or, e.g., through the hierarchy. To highlight the concept, the following definition
is constructed.

\textbf{Definition.} We say that an intelligent system is embodied in an intelligent, possibly hierarchical
environment, if (a) it can learn and solve combinatorial problems, (b) can communicate the solutions, and (c) if it can
be engaged in (distributed) verification of solutions communicated by other intelligent agents.

To build a network model, verification is identified as the opposite of encoding. Verification of an encoded quantity
means (i) decoding, i.e., the reconstruction of inputs using communicated encoded quantities, (ii) comparison of the
reconstructed input and the real input. In turn, a top-down model of neocortical processing of sensory information can
make use of generative models equipped with comparators, in which the distributed hierarchical decoding process is to
be reinforced by comparisons of the input and the decoded quantities.  This shall be our computational model for CoI.

\section{Model description}

\begin{figure}[h]
\centering
   \includegraphics[width=93mm]{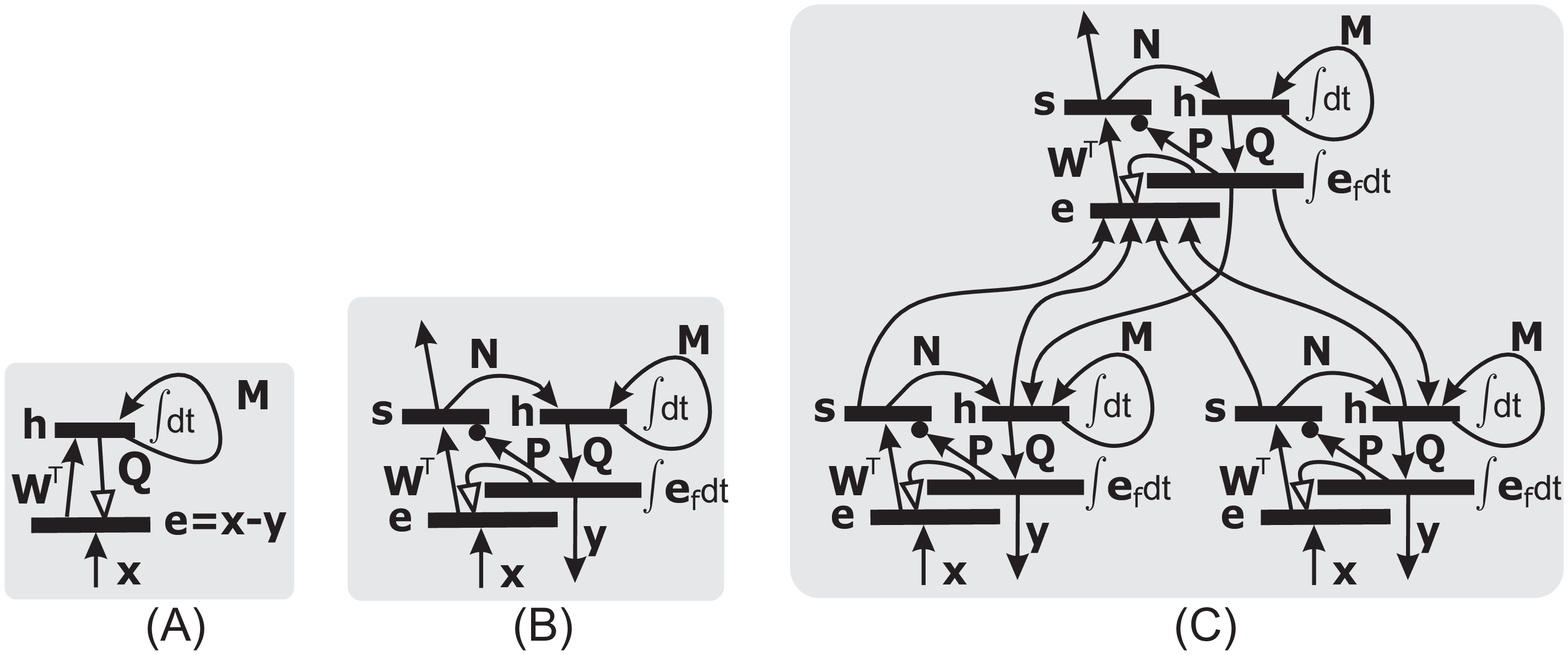}
   \caption{\textbf{Reconstruction networks}
\newline   \noindent  \textbf{A:} Simple reconstruction network (RCN).  \textbf{B:} RCN with
sparse code shrinkage noise filtering and non-negative matrix factorization. \textbf{C:} RCN hierarchy.
  (See text.) }\label{f:recnets}
\end{figure}

Encoding, decoding and comparison is performed by reconstruction networks. The basic \textit{reconstruction loop}
(Fig.~\ref{f:recnets}A) has two layers: the reconstruction error layer that computes the difference (${\bf e} \in
\mathrm{R}^r$) between input (${\bf x} \in \mathrm{R}^r$) and reconstructed input (${\bf y} \in \mathrm{R}^r$): ${\bf
e}={\bf x}-{\bf y}$ and the \textit{hidden} layer that holds the hidden \textit{internal representation} ${\bf h} \in
\mathrm{R}^s$ and produces the reconstructed input ${\bf y}$ via top-down transformation ${\bf Q} \in \mathrm{R}^{r
\times s}$. The hidden representation is corrected by the bottom-up transformed form of the reconstruction error ${\bf
e}$, i.e., by ${\bf We}$, where ${\bf W}\in \mathrm{R}^{s \times r}$ and is of rank $\min (s,r)$. The process of
correction means that the previous value of the hidden representation is to be maintained and the correcting amount
needs to be added. In turn, the hidden representation has self-excitatory connections (${\bf M}$), which sustain the
activities. For sustained input ${\bf x}$, the iteration will stop when ${\bf WQh}={\bf Wx}$: The relaxed hidden
representation is \textit{solely} determined by the input and top-down matrix ${\bf Q}$. The latter is identified with
the long-term memory. BU matrix ${\bf W}$ is perfectly tuned if ${\bf W}=({\bf Q}^T{\bf Q})^{-1}{\bf Q}^T$, i.e., if
${\bf WQ}={\bf I}$ (superscript $T$ denotes transposition and ${\bf I} \in \mathrm{R}^{s \times s}$). In this case, the
network acts and is as fast as a feedforward net.

The network can be extended to support noise filtering by (i) separating the reconstruction error layer and the
reconstructed input layer, (ii) adding another extra layer that holds ${\bf s}={\bf W}{\bf e}$ and transformation ${\bf
s}\rightarrow{\bf h}$, i.e., ${\bf N}$ supports pattern completion \cite{lorincz02mystery} and (iii) assuming that BU
transformation ${\bf W}$ maximizes BU information transfer by minimizing mutual information (MMI) between its
components. BU transformed error is passed to the hidden representation layer through transformation matrix ${\bf N}$
and corrects hidden representation ${\bf h}$. Learning of the hidden representation may make use of e.g., non-negative
matrix factorization \cite{lee99learning} (Fig.~\ref{f:recnets}(B)) or an entropic prior on WTA subnets
\cite{szatmary04accepted}.

MMI plays an important role in noise filtering. There are two different sets of afferents to the MMI layer: one carries
the error, whereas the other carries the reconstructed input ${\bf y}$ via bottom-up transformation ${\bf P}$, followed
by a non-linearity that removes noise via thresholding. Thresholding is alike to wavelet denoising, but filters are not
necessarily wavelets: They are optimized for the input database experienced by the network. MMI algorithms enable the
local estimation and local thresholding of noise components. The method is called sparse code shrinkage (SCS)
\cite{hyvarinen99sparse2}. Note that SCS concerns the components of the BU transformed reconstructed input: high (low)
amplitude components of the BU transformed reconstructed input ${\bf Py}$ can (not) open the gates of components of the
MMI layer and MMI transformed reconstruction error can (not) pass the open (closed) gates to correct the hidden
representation. Apart from SCS, the reconstruction network is linear. We shall denote this property by the sign
`$\sim$'. `$A$ $\sim$ $B$' means that up to a scaling matrix, quantity $A$ is approximately equal to quantity $B$. For
a \textit{well tuned network }and if matrix ${\bf M}$ performs temporal integration, then ${\bf \dot{x}}\cong {\bf e}
\sim{\bf s}$ by construction. In a similar vein, hidden representation ${\bf h} \sim {\bf y}$, ${\bf y} = \int {\bf
e}_f dt$ where ${\bf e}_f$ is the noise filtered version of ${\bf e}$, and, \textit{apart from noise}, ${\bf y}$
approximates ${\bf x}$. In a hierarchy of such networks, beyond other terms, ${\bf h}$ can be transmitted to higher
networks and the higher networks may overwrite the internal representation of the lower ones. Non-linear effects are
introduced by SCS and by the top-down overwriting of internal representation (Fig.~\ref{f:recnets}C). This latter -- by
construction -- is equivalent to denoised direct communication of correlations between internal representations.

\section{Discussion}

The reconstruction (also called generative) network concept provides a straightforward resolution to the homunculus
fallacy (see, e.g., \cite{searle92rediscovery}). The fallacy says that no internal representation is meaningful without
an interpreter, `who' could `make sense' of the representation. Unfortunately, all levels of abstraction require at
least one further level to become the corresponding interpreter. Thus, interpretation is just a new transformation and
we are trapped in an endless regression.

\begin{figure}[h!]
\centering
   \includegraphics[width=6cm]{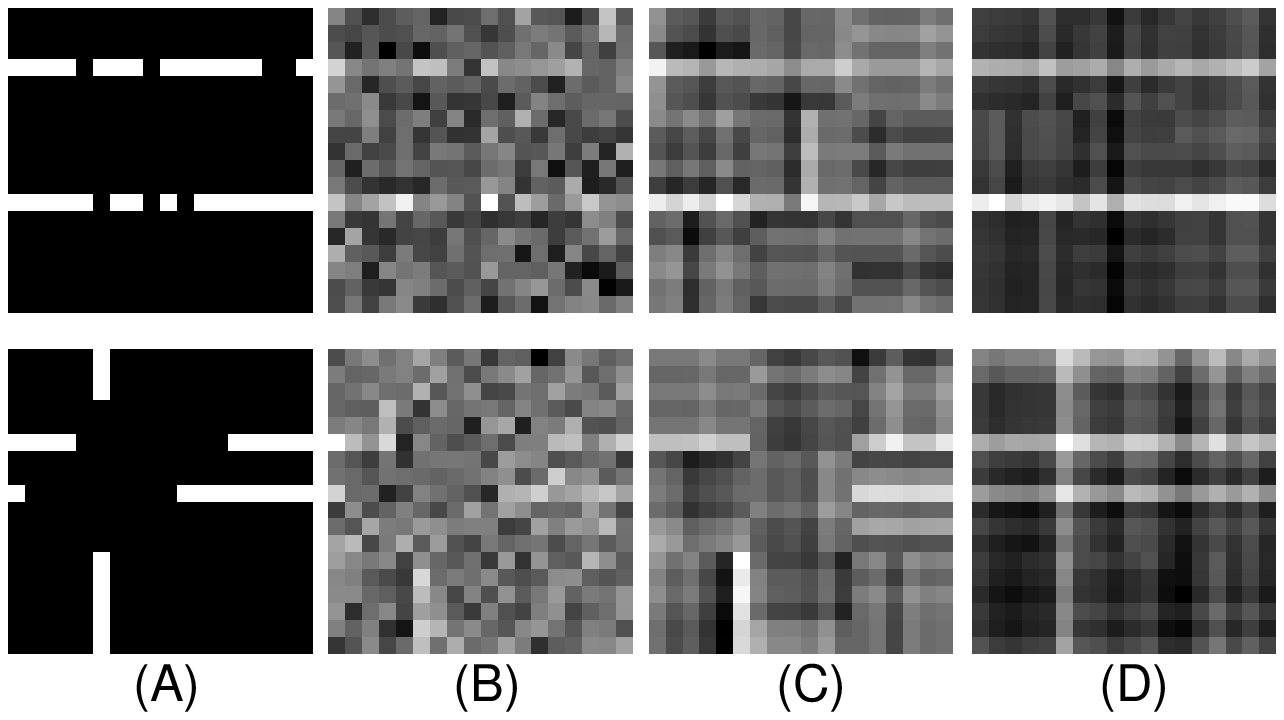}
   \caption{\textbf{Improved pattern completion in the hierarchy.}\newline
  Subnets of the lower layer have input dimension of $6 \times 6 =36$.
  Lower layer is made of $3 \times 3 =9$ subnets with $\dim(\mathbf{h}_{lower})=12$, each.
  $\dim(\mathbf{x}_{higher})=9\times 12=108$. $\dim(\mathbf{h}_{higher})=36$.
  \textit{Upper row:} Pixels of the inputs are missing. \textit{Bottom row:}
  Pixels \textit{and} sub-components of the inputs are missing.
  (A): Noiseless input.
  (B): Input applied (during learning and working).
  (C): Lower layer reconstructions.
  (D) Reconstruction for the full hierarchy with top-down overwriting of the internal representations.
}\label{f:hier_results}
\end{figure}

Reconstruction networks turn the fallacy upside down by changing the roles \cite{lorincz97towards}: Not the internal
representation but the \textit{input} `makes sense', if the same (or similar) inputs have been experienced before and
if the input can be derived/generated by means of the internal representation. In reconstruction networks, infinite
regression occurs in a finite loop and progresses through iterative corrections, which converge. Then the fallacy
disappears. In our wording, (i) the internal representation interprets the input by (re)constructing components of the
input using the components of the internal representation and that (ii) component based generative pattern completion
`makes sense' of the input. We shall illustrate the idea by reviewing computational simulations on a combinatorial
problem depicted in Fig.~\ref{f:hier_results} \cite{lorincz02mystery}.

\noindent In these simulations, several horizontal and vertical bars were presented to the network of
Fig.~\ref{f:recnets}C. Neurons of subnetworks did accomplish the combinatorial tasks and separated the components, the
horizontal and vertical bars as revealed by the reconstructed inputs in Fig.~\ref{f:hier_results}C. (See also
\cite{lorincz02mystery}). The reconstruction network higher in the hierarchy collected information from the lower
networks and overwrote the internal representation of the lower networks. The encoded information \textit{collected
from} and \textit{communicated to} lower networks improved pattern completion (Fig.~\ref{f:hier_results}D).

In our formulation, `linguistic theft' occurs, e.g., if internal representations are `stolen' from a `toiler' for
(self-)supervised training of the `thief' applying Hebbian learning, e.g., for noiseless input $\mathbf{x}_{noisefree}$
and the stolen representation $\mathbf{h}_{stolen}$: $\Delta \mathbf{Q} \propto
\mathbf{x}_{noisefree}\mathbf{h}_{stolen}^T$. Recall that during processing, the new internal representation shall be
determined by input $\mathbf{x}$ and the new $\mathbf{Q}$. The immediate improvement of pattern completion abilities
`justifies' the `theft' and trial-by-error learning can be dismissed. Moreover, exponential gains can be achieved for
WTC type problems. In turn, reconstruction networks equipped with Hebbian learning suit `linguistic theft' for WTC type
problems. Similar gains can be achieved if the information processed by subnetworks overlap and if the networks can
transfer representations to each other.

\subsection{Connections to neuroscience}

It has been demonstrated \cite{lorincz02mystery} that maximization of information transfer is an emerging constraint in
reconstruction networks. Here, we note that the model provides straightforward explanation for the differences found
between neurons of the deep and superficial layers of the entorhinal cortex (i.e., that deep layer neurons have
sustained responses, whereas superficial layer neurons do not) \cite{egorov02graded}, which is the consequence of the
sustained activities in the hidden layer. The model also explains the long and adaptive delays found recently in the
dentate gyrus \cite{henze02single}, which -- according to the model -- should be there but are not necessary anywhere
else along the visual processing stream. Last but not least, the model makes falsifying predictions about the feedback
connections between visual processing areas, which -- according to the mapping to neocortical regions -- correspond to
the long-term memory of the model.

\section{Conclusions}

Our goal was to underpin a recent model of neocortical information processing \cite{lorincz02mystery} by means of the
theory a computational complexity. We have argued that sensory processing areas developed by evolution can be viewed as
intelligent agents using economical communication. According to our argument, the agents encode solutions to
combinatorial problems (NP-hard problems, or `components' in terms of psychology \cite{biederman87recognition}),
communicate the encoded information and decode the communicated information. We have argued that reconstruction
networks equipped with Hebbian learning are appealing for `linguistic theft' of solutions of such problems. We have
reviewed computational experiments where a combinatorial component-search problem was (1) solved and (2) the
communicated encoded information was decoded and used to improve pattern completion. The novelty of the present work is
in the reinterpretation of a recent model of neocortical information processing in terms of communicating agents, who
(i) communicate encoded information about combinatorial learning tasks and (ii) cooperate in input verification by
decoding the received and encoded quantities. We have also described the straightforward predictions of the model that
have relevance to visual neuroscience.


\bibliographystyle{apalike}
\small{

}

\end{document}